\documentclass[11pt]{article}
\usepackage{moriond,epsfig}
\begin{document}
\vspace*{4cm}
\title{Hadronic Structure from Perturbative Dressing in QCD - The Valon Model}

\author{ Firooz Arash$^{(a)}$\footnote{e-mail: farash@cic.aut.ac.ir} and Fatemeh Taghavi$^{(b)}$ }

\address{
$^{(a)}$ Physics Department, Tafresh University, Tafresh, Iran \\
$^{(b)}$ Physics Department, Iran University of Science and
Technology, Narmak, Tehran, Iran \\
} \maketitle\abstracts{In the Framework of {\it{valon model}} we
have calculated the parton distribution in a valon for both the
polarized and unpolarized cases. These distributions are
originated purely from the dressing of a valence quark in QCD and
are common to all hadrons. Structure functions, $F_{2}^{p,\pi}$,
are obtained, which agree rather well with the experimental
results. A simple relation between $F_{2}^{p}$ and $F_{2}^{\pi}$
is inferred. For the polarized structure function, $g_{1}$ while
the model gives good agreement with data, it requires a sizable
angular momentum contribution to the spin of the valon, and hence
to that of proton. This contribution is calculated.  }
{\bf{Introduction}} The parton distributions in proton have been
studied extensively in a wide range of both $x$ and $Q^{2}$. The
parton distribution functions (PDF) are calculated numerically by
fitting over 1300 data points which rely on over 30 parameters in
the input PDF's. Although these PDF's are accurate but
inconvenient to describe analytically. Recently, some experimental
results have also emerged for the pion \cite{1}, at very low $x$
which can serve as a further check on our understanding of hadron
structure. Here we elaborate on the valon model that can be very
useful in the study of hadronic structure, in particular when the
experimental data are scarce. A valon has its own cloud of partons
which can be calculated in pQCD. This structure is universal and
independent of the hosting hadron. In \cite{2} the structure of
the valon is calculated to the Next-to-leading Order in QCD and
tested against the $F_{2}^{p} (x, Q^{2})$ in a wide range of
kinematics: $x=[10^{-4},1]$ and $Q^{2}=[1, 5000]$ $GeV^2$. Here we
will examine its ability in describing pion structure functions,
$F_{2}^{\pi}$ without any additional new parameter. This is
important, because at first it is not clear at all that if one
moves away from proton, where there are ample data points, to pion
where the data are much restricted, the insight gained from the
study of proton, will still be valid. At the end we will discuss
the case of polarized structure function $g_{1}$.\\
{\bf{Formalism: Unpolarized}} The unpolarized structure of a valon
is calculated in next-to-leading order framework of PQCD in
Ref.[2]. Without going into the details it suffice here to give
the parameterized form of the valonic PDF as follows:
\begin{equation}
zq^{valon}_{valence}(z,Q^{2})=a z^{b}(1-z)^{c},
\end{equation}
\begin{equation}
zq^{valon}_{sea}(z,Q^{2}) = \alpha z^{\beta}(1-z)^{\gamma}[1+\eta
z +\xi z^{0.5}].
\end{equation}
The parameters $a$, $b$, $c$, $\alpha$, $\beta$, $\gamma$, $\eta$,
and $\xi$ are functions of $Q^{2}$ and are given in the appendix
of Ref. [2]. Gluon distribution in a valon has an identical form
as in Eq. (2) but with different parameter values. Upon
convolution of these valonic PDF's with the valon distribution,
$G_{valon}^{h}(y)$ in a hadron, we obtain the hadronic structure
function:
\begin{equation}
F_{2}^{p}(x,Q^{2})=
 \sum_{valon} \int_{x}^{1}dy
G_{valon}^{h}(y) F_{2}^{valon}((\frac{x}{y}), Q^{2})
\end{equation}
where $G_{valon}^{h}(y)$ is the distribution of valon with
momentum fraction $y$ in hadron h. These functions are given in
Refs.\cite{2} \cite{3} \cite{4} for proton, pion, and kaon.
$F_{2}^{valon}$ is the structure function of a valon. In figures
(9) of Ref.[2] the results for the proton structure function is
presented at $1<Q^2 <2000$ $GeV^2$. They are compared with the
experimental values and with other global fitting results, the
agreements are excellent. For the details see Ref.[2].\\
Pion structure function at low $x$ in the leading neutron
experiment [1] is reported. Moreover, there are some data from
Drell-yan \cite{5} experiments at medium and large $x$ for the
valence quark distribution. We have used the data of Ref.[5] at
$Q^2=25 GeV^2$ to determine$G_{valon}^{\pi}(y)$. In Figure [1] the
results are shown along with the SMRS and  GRS determination. It
turns out that the valon distribution in pion is a very broad:
\begin {equation}
G_{\frac{U}{\pi^+}(y)}=G_{\frac{\bar{D}}{\pi^+}(y)}=1.02
(1-y)^{0.01} y^{0.01}
\end{equation}
This is expected; for it indicates that the valons in a pion are
tightly bound, a reflection of the fact that the pion is much
lighter than the mass of its valons. From Eqs.(1-5) $F_{2}^{\pi}$
is readily calculated without any free parameters. In Figure (2)
we present $F_{2}^{\pi}$at $Q^2=7$ and $Q^2=15$ $GeV^2$ at very
low $x$ along with the data from Ref.[1]and the results from SMRS
and GRS. In the figure there are two sets of data points,
corresponding to two different normalization of the data by ZEUS
coll. They differ by a factor of two [1]. Our results favor the
effective flux normalization. Same is true for SMRS , but GRS is
more compatible with the additive quark model normalization.
However, as can be seen from figure (1) GRS determination grossly
disagrees with the valence quark distribution in pion. In \cite{3}
and \cite{4} it is shown that the valon description holds true for
all values of $Q^2$.
\begin{figure}[htb]
\vspace{2pt}
\includegraphics[angle=0,width=18pc,height=8pc]{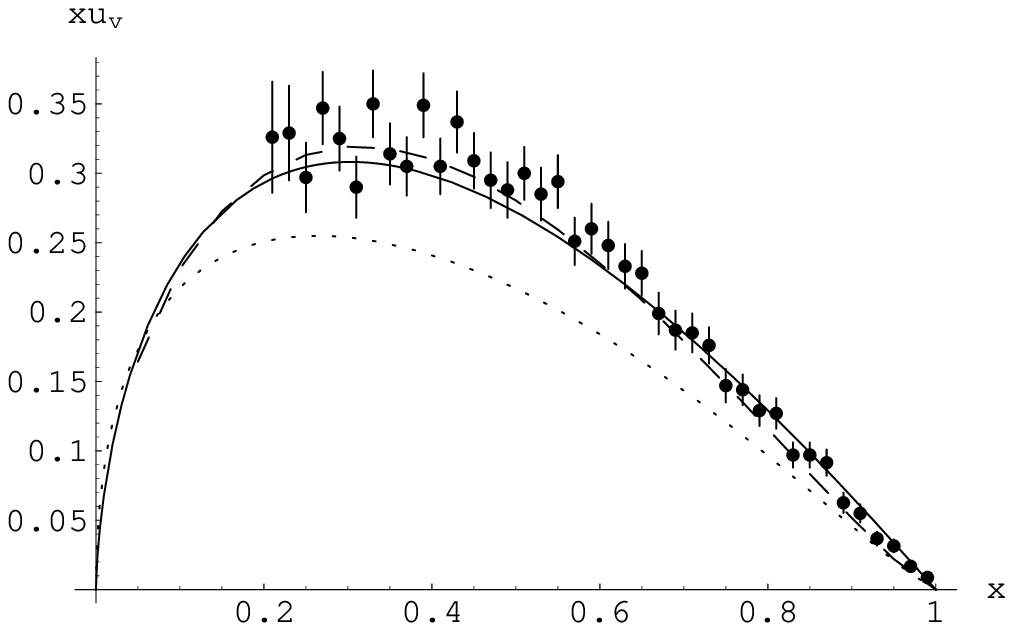}
\includegraphics[angle=0,width=18pc,height=10pc]{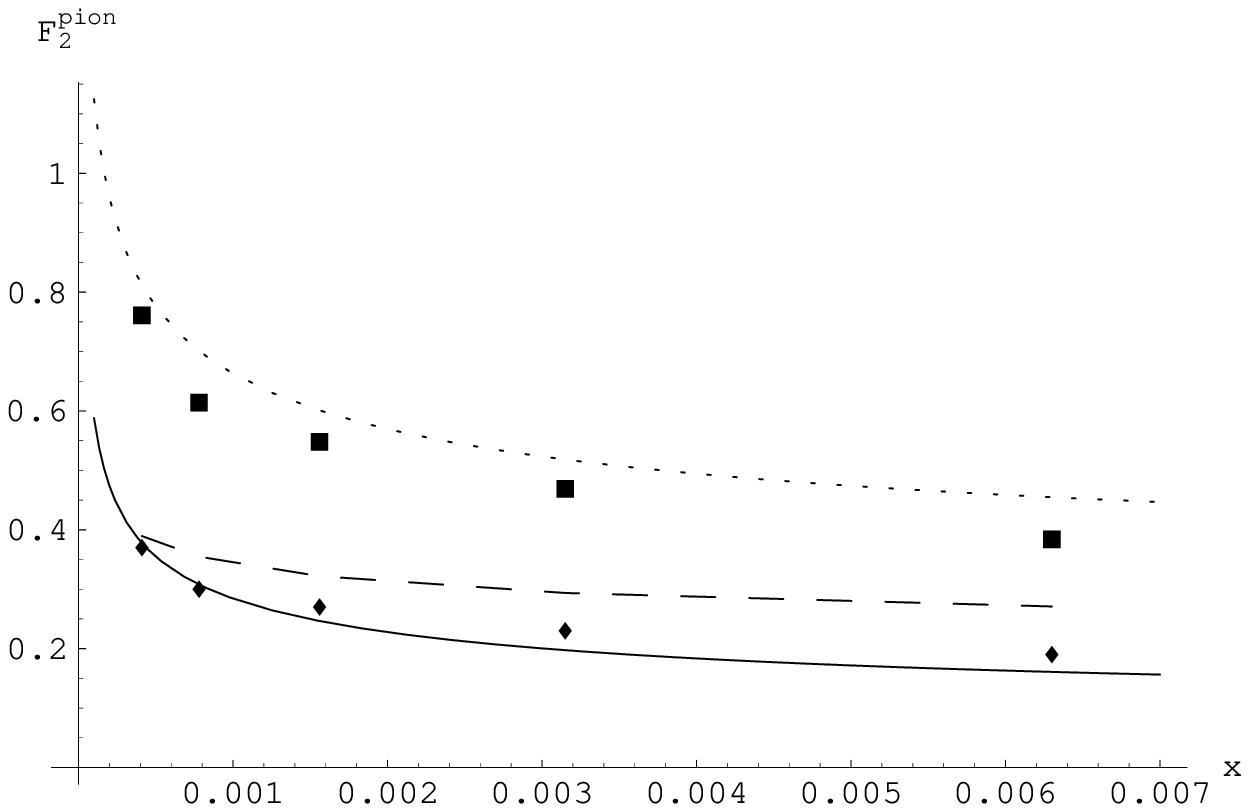}
\caption{\footnotesize {\bf{Right}}- Comparison of the pion
valence distribution, $u_{valence}^{\pi^{-}}(x)$, using the valon
model calculation (solid line), SMRS (dashed line),and GRS (dotted
line), with the data from Ref.[1] at $Q^{2}=25$ $GeV^{2}$.\newline
{\bf{Left}}- Pion structure function, $F_{2}^{\pi}(x)$, at $Q^2=7$
$GeV^{2}$. The diamonds and the squares are pion flux and additive
quark model normalization of the data [1], respectively. The solid
line is the model calculation; The dashed line is from SMRS and
the dotted line is from GRS determination. }
\label{figure 1.}
\end{figure}

\begin{figure}[htb]
\vspace{2pt}
\includegraphics[angle=0,width=18pc,height=10pc]{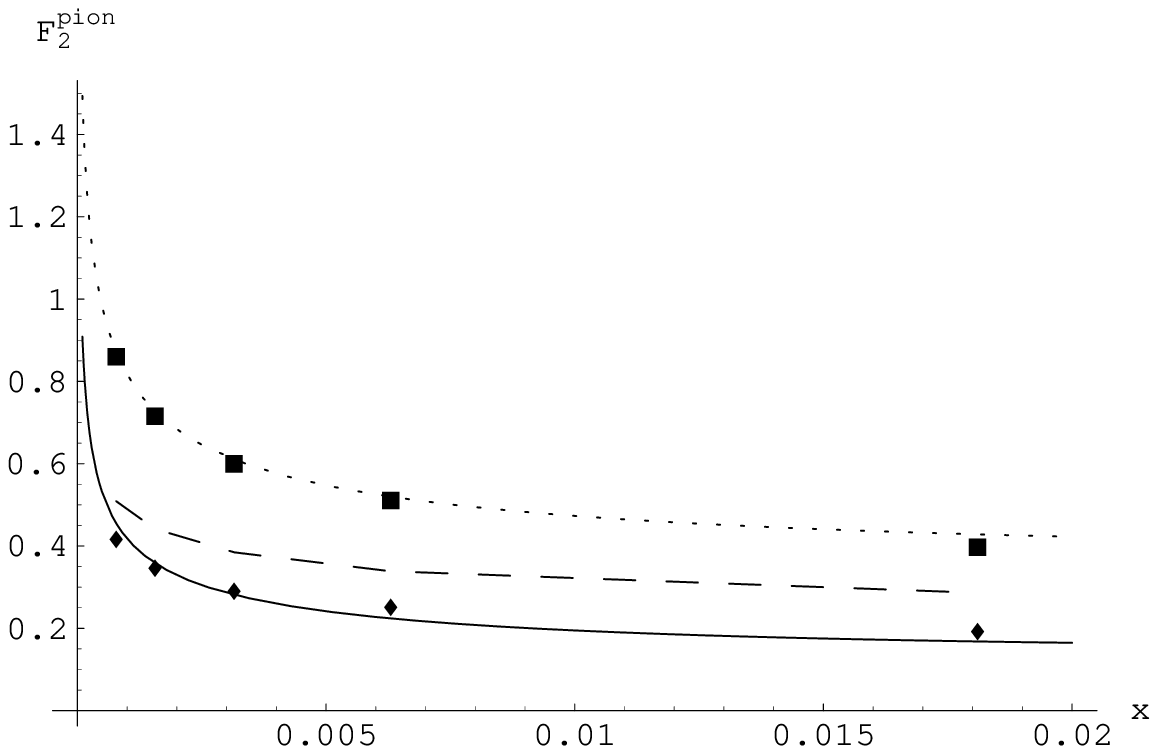}
\includegraphics[angle=0,width=18pc,height=10pc]{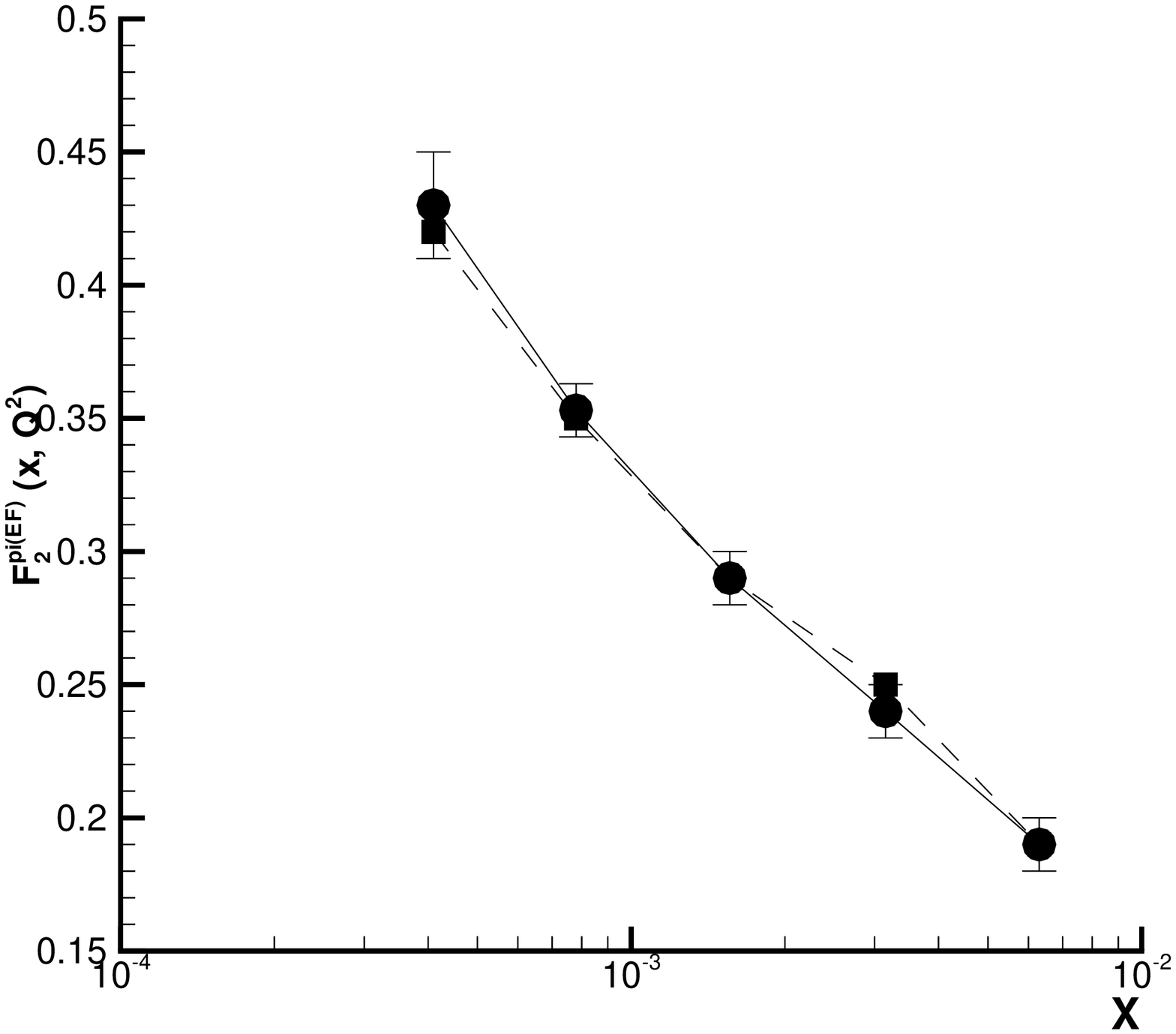}
\caption{\footnotesize {\bf{Right}} The same as the left figure in
Figure-1 but at $Q^2=15$ $GeV^2$. {\bf{Left}}- The effective flux
normalization of $F_2^{\pi}$ data as compared with the scaled
$F_2^{p}$ at$Q^2=15 GeV^2$ of equations (5). The data points are
from Ref.[1]. The solid line is $0.371 F_2^{p}(x,Q^2)$ pertinent
to Eq. (5) and the dashed line is $F_2^{\pi}$, both are calculated
directly and separately from the model. } \label{figure 2.}
\end{figure}
It is also interesting to notice that a simple relationship holds
between $F_{2}^{p}$ and effective flux normalization of pion
structure function $F_{2}^{\pi(EF)}$ which is also notices by ZEUS
Coll. in Ref. [1]; namely:
\begin{equation}
F_{2}^{\pi(EF)}(x,Q^2)\approx kF_{2}^{p}(x,Q^2)
\end{equation}
With $k\approx 0.37$. We have calculated both sides of equations
(6) independently, using the valon model; a sample result at
$Q^2=15$ $GeV^2$ is given in Figure (3), but the relationships
also hold true for other values of $Q^2$. \\
{\bf{Polarized Structure Functions}} The case of polarized
structure function requires knowledge of polarized parton
distribution in a polarized valon and the distribution of
polarized valon in polarized hadron. We will restrict ourselves to
nucleon. The first task is carried out similar to that of
unpolarized case; namely, the moments of polarized structure
functions in a valon is calculated in QCD , then with Inverse
Mellin transform technique the polarized parton densities are
obtained. As for the second requirement, we define the function,
$\delta F_{j}(y)$ via:
\begin{equation}
\delta G_{j} (y)=\delta F_{j}(y) G_{j}(y)
\end{equation}
where, $j$ refers to U or D valon, and $G_{J}(y)$ is the
unpolarized valon distribution. In addition, we impose upon
$\delta F_{j}(y)$, the constraint that $P_{U}=2/3$ and $P_{D}=
-1/3$, corresponding to $SU(6)$ model, with  $P_{j}$ defined by
$P_{j}$=$\int_{0}^{1} dy \delta F_{j}(y)G_{J}(y)$. Then, for a
nucleon we have:
\begin{eqnarray}
g_{1}^{N}(x,Q^2)=\int_{x}^{1} dy[ 2 \Delta G_{U}g_{1}^{U}
(\frac{x}{y},Q^2)+  \Delta G_{D}g_{1}^{UD} (\frac{x}{y},Q^2)]
\end{eqnarray}
This completes the formalism. In Figure (4) $xg_{1}^{p}$ are
shown. It is evident that the model is able to reproduce the
experimental data.
\begin{figure}[htb]
\vspace{2pt}
\includegraphics[angle=0,width=18pc,height=16pc]{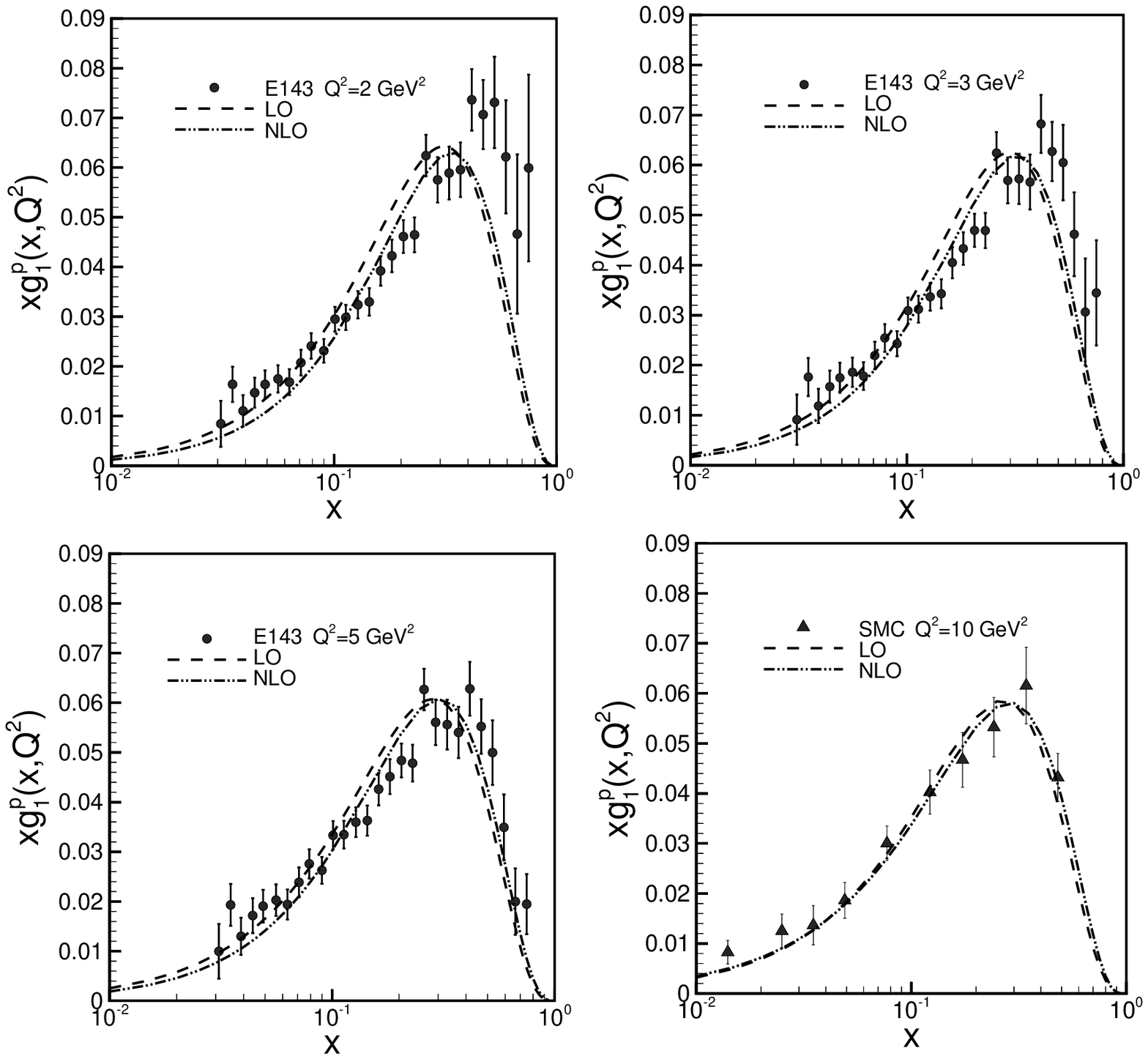}
\includegraphics[angle=0,width=18pc,height=16pc]{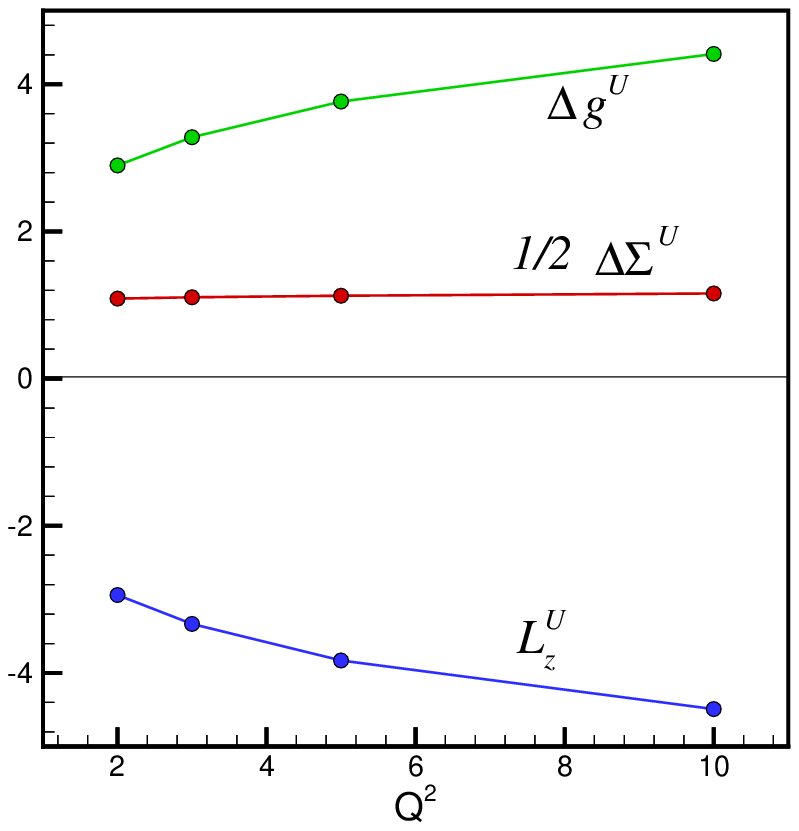}
\caption{\footnotesize{\bf{Right}}- $xg_{1}^{p}(x,Q^2)$ as
calculated from the model and compared with the experimental
data.\hspace{2.5cm} {\bf{Left}}- various component contributions
to the spin of a valon as a function of $Q^2$ } \label{figure 3.}
\end{figure}
Much interest has been centered on the integral
\begin{equation}
\Gamma^{p(n)}=\int dx g_{1}^{p(n)}(x).
\end{equation}
Our model gives $\Gamma^{p}=0.1212, 0.1219,0.1225$, and $0.1236$
for $Q^2=2,3,5,10$ $GeV^2$, respectively, which are very close to
the experimental data. The same is true for $\Gamma^{n}$. These
results suggest that quark contribution to the spin of nucleon is
small. So, while the model produces all available data on
polarized structure function, but the same data do not give the
nucleon spin. Therefore, it is natural to ask if the spin content
of a valon do give the spin $\frac{1}{2}$ of the valon itself. The
contribution of different components of a valon to its spin is
obtained by the following integral, say for a U-type valon.
\begin{equation}
\Delta H_{\frac{i}{U}}(Q^2)=\int_{0}^{1} \delta
q_{\frac{i}{U}}(z,Q^2) dz
\end{equation}
where $i$ is for valence, sea, and gluon inside a valon. For the
valence part, $\Delta H_{\frac{valence}{U}}$ is constant for all
$Q^2$ values being equal to one as one would expect.$\Delta
H_{\frac{sea}{U}}$ while varying with $Q^2$, remains small around
$0.08-0.2$ for a range of $Q^2=2-10$ $GeV^2$.  However $\Delta
H_{\frac{gluon}{U}}$ is fairly large and grows rapidly with $Q^2$,
reaching to 4.4 at $Q^2=10$ $GeV^2$. Such a large value, of
course, is not unexpected according to QCD evolution equation.
Thus, by considering only the numerical values of  eq. (13), it is
impossible to build a spin $\frac{1}{2}$ valon without an extra
element.  It turns out that the orbital angular momentum of sea
partons in a valon is the extra element in the following sum rule
\begin{eqnarray}
S^{U}_{z}=\frac{1}{2}(S^{valence}_{z} + S^{sea}_{z})^U
+(S^{gluon}_{z})^U + L_{z}^U= \frac{1}{2}
\end{eqnarray}
It is known that there is a mixing operator which dresses any
valence quark wave function with $\bar{q}-q$ pares to respect
PCAC. The vacuum of these massive valons is a coherent
superposition of {\it {cooper}} pairs of massless quark, which
resembles the structure of a valon to that of superconductivity.
Studies of anistropic superconductivity shows that a particle of
condensed fluid surrounded by a cloud of correlated particles must
rotate around it with the axis of rotation {\it{l}}. The presence
of anistropy leads to axial asymmetry of pairing correlations
around the anistropy direction. The axis of asymmetry can be
associated with the polarization vector of the valence quark
located at the center of the cloud. It can be shown that the
magnitude of this $l_{z}$ associated with the so-called {\it{back
flow}} must be negative, which is precisely what we need to
compensate the growth of gluon polarization contribution. In
figure (5) this fact is depicted.

\end{document}